\documentclass[a4paper,12pt]{article}
\usepackage{Stylefile}
\title
{ %
\vspace*{3.0cm} \LARGE{\bf Modeling the Flow of Yield-Stress Fluids in Porous Media} \vspace*{4.0cm} \\
}

\author{Taha Sochi\footnote{University College London - Department of Physics \& Astronomy - Gower Street - London. Email:
t.sochi@ucl.ac.uk.} \vspace*{5.0cm}}

\begin{document}

\maketitle %
\pagenumbering{arabic}

\newpage
\phantomsection \addcontentsline{toc}{section}{Contents} %
\tableofcontents

\phantomsection \addcontentsline{toc}{section}{List of Figures} %
\listoffigures

\newpage
\phantomsection \addcontentsline{toc}{section}{Abstract} \noindent
{\noindent \LARGE \bf Abstract} \vspace{0.5cm}\\
\noindent %

Yield-stress is a problematic and controversial \nNEW\ flow phenomenon. In this article, we
investigate the flow of yield-stress substances through porous media within the framework of
pore-scale network modeling. We also investigate the validity of the Minimum Threshold Path (MTP)
algorithms to predict the pressure yield point of a network depicting random or regular porous
media. Percolation theory as a basis for predicting the yield point of a network is briefly
presented and assessed. In the course of this study, a yield-stress flow simulation model alongside
several numerical algorithms related to yield-stress in porous media were developed, implemented
and assessed. The general conclusion is that modeling the flow of yield-stress fluids in porous
media is too difficult and problematic. More fundamental modeling strategies are required to tackle
this problem in the future.

%%%%%%%%%%%%%%%%%%%%%%%%%%%%%%%%%%%  Head style  %%%%%%%%%%%%%%%%%%%%%%%%%%%%%%%%%%%
\pagestyle{headings} %
\addtolength{\headheight}{+1.6pt}
\lhead[{Chapter \thechapter \thepage}]%
      {{\bfseries\rightmark}}
\rhead[{\bfseries\leftmark}]%
     {{\bfseries\thepage}} %tell it to put page number at rhead
\headsep = 1.0cm               % Added 07 Sep 2006
%%%%%%%%%%%%%%%%%%%%%%%%%%%%%%%%%%%%%%%%%%%%%%%%%%%%%%%%%%%%%%%%%%%%%%%%%%%%%%%%%%%%

\newpage
%XXXXXXXXXXXXXXXXXXXXXXXXXXXXXXXXXXXXXXXXXXXXXXXXXXXXXXXXXXXXXXXXX
\section{Introduction} \label{Introduction}

\Yields\ or viscoplastic fluids are characterized by their ability to sustain shear stresses, that
is a certain amount of stress must be exceeded before the flow initiates. So an ideal \yields\
fluid is a solid before yield and a fluid after. Accordingly, the viscosity of the substance
changes from an infinite to a finite value. The physical situation, however, indicates that it is
more realistic to regard \yields\ substances as fluids whose viscosity as a function of applied
stress has a discontinuity as it drops sharply from a very high value on exceeding a critical
stress.

There are many controversies and unsettled issues in the \nNEW\ literature about \yields\
phenomenon and \yields\ fluids. In fact, even the concept of a \yields\ has received much recent
criticism, with evidence presented to suggest that most materials weakly yield or creep near zero
strain rate. The supporting argument is that any material will flow provided that one waits long
enough. These conceptual difficulties are backed by practical and experimental complications. For
example, the value of \yields\ for a particular fluid is difficult to measure consistently and it
may vary by more than an order of magnitude depending on the measurement technique \cite{birdbook,
carreaubook, Barnes1999, BalmforthC2001}.

Among the difficulties in working with \yields\ fluids and validating the experimental data is that
\yields\ value is usually obtained by extrapolating a plot of shear stress to zero shear rate
\cite{carreaubook}. This can result in a variety of values for \yields, depending on the distance
from the shear stress axis experimentally accessible by the instrument used. The vast majority of
\yields\ data reported results from such extrapolations \cite{ParkHB1973, alfariss1}, making most
values in the literature instrument-dependent \cite{carreaubook}. Another method used to measure
\yields\ is by lowering the shear rate until the shear stress approaches a constant. This may be
described as a dynamic \yields\ \cite{larsonbook1999}. The results obtained using such methods may
not agree with the static \yields\ measured directly without disturbing the microstructure during
the measurement. The latter seems more relevant to the flow initiation under gradual increase in
pressure gradient as in the case of porous media flow experiments. Consequently, the accuracy of
the predictions made using flow simulation models in conjunction with such experimental data is
limited.

Another difficulty is that while in the case of pipe flow the \yields\ value is a property of the
fluid, in the case of flow in porous media it may depend on both the fluid and the porous medium
itself \cite{bearbook, vradis1}. One possible explanation is that \yields\ value may depend on the
size and shape of the pore space when the polymer macromolecules become comparable in size to the
pore. The implicit assumption that \yields\ value at pore level is the same as the value at bulk
may not be evident. This makes the predictions of the models based on analytical solution to the
flow in a uniformly-shaped tube combined with the bulk rheology less accurate. When the duct size
is small, as it is usually the case in porous media, flow of macromolecule solutions normally
displays deviations from predictions based on corresponding viscometric data \cite{LiuM1998}.
Moreover, the highly complex shape of flow paths in porous media must have a strong impact on the
actual yield point, and this feature is lost by modeling these paths with ducts of idealized
geometry.

Various attempts to model the flow of yield-stress fluids in porous media can be found in the
literature. Most of these are based on extending the existing continuum non-yield-stress flow
models to incorporate yield-stress. For example, Pascal \cite{pascal1} modified \Darcy's law by
introducing a threshold pressure gradient to account for the \yields. This threshold gradient is
directly proportional to the \yields\ and inversely proportional to the square root of the absolute
permeability. Vradis and Protopapas \cite{vradis1} extended the `capillary tube' and the
`resistance to flow' models to describe the flow of \BING\ fluids in porous media and presented a
solution in which the flow is zero below a threshold head gradient and Darcian above it. Chase and
Dachavijit \cite{chase1} modified the \Ergun\ equation to describe the flow of \yields\ fluids
through porous media by applying a bundle of capillary tubes approach. Other attempts include Wu
\etal\ \cite{wu1} who used an integral analytical method to obtain an approximate analytical
solution for single-phase flow of \BING\ fluids through porous media. Chaplain \etal\
\cite{chaplain1} also modeled the flow of \BING\ fluids through porous media by generalizing
Saffman \cite{saffman1} analysis for the \NEW\ flow to describe the dispersion in a porous medium
by a random walk. Recently, Balhoff and Thompson \cite{balhoff1} used their three-dimensional
network model, which is based on a computer-generated random sphere packing, to investigate the
flow of \BING\ fluids in packed beds. Network modeling was also used by several researchers to
investigate the generation and mobilization of foams and \yields\ fluids in porous media. These
include Rossen and Mamun \cite{rossen2} and Chen \etal\ \cite{ChenYR2004, ChenYR2005, chen3}.

\newpage
%SSSSSSSSSSSSSSSSSSSSSSSSSSSSSSSSSSSSS
\section{Modeling \YieldS\ in Porous Media} \label{ModelingYS}

Several constitutive equations to describe the fluids with \yields\ are in use; the most popular
ones are Bingham, Casson and \HB. Some have suggested that the \yields\ values obtained via such
models should be considered model parameters and not real material properties.

There are four main approaches for modeling the flow through porous media in general. These are:
continuum models, bundle of tubes models, numerical methods and pore-scale network modeling. It
seems that all these approaches suffer from difficulties when applied to the flow of yield-stress
fluids in complex porous media.

The failure of the continuum models should not be a surprise because these models do not account
for the complex geometry and topology of the void space. As the yield point depends on the fine
details of the pore space structure, no continuum model is expected to predict the threshold yield
pressure. The continuum models also fail to predict the flow rate, at least at transition stage
where the medium is partly conducting, because according to these models the medium is either fully
blocked or fully conducting whereas in reality the yield of porous medium is a gradual and
pressure-dependent process. For instance, these models predict for \BING\ fluids a linear
relationship between Darcy velocity and pressure gradient with an intercept at the threshold yield
gradient whereas network modeling results, supported by experimental evidence, predict a nonlinear
behavior at transition stage \cite{vradis1}. Some authors have concluded that in a certain range
the macroscopic flow rate of \BING\ plastic in a network depends quadratically on the departure of
the applied pressure difference from its minimum value \cite{chen3}. Our network simulation
predictions confirm this quadratic correlation.

Regarding the capillary bundle models, the situation is similar to the continuum models as they
predict a single universal yield point at a particular pressure drop if a uniform bundle of
capillaries is assumed, whereas in reality porous medium yield occurs gradually as the pressure
gradient increases. Moreover, because all capillary bundle models fail to capture the topology and
geometry of complex porous media, they cannot predict the yield point and describe the flow rate
since yield is highly dependent on the fine details of the void space. An important aspect of the
geometry of real porous media, which strongly affects the yield point and flow rate of \yields\
fluids, is the \convdiv\ nature of the flow paths. This feature is not reflected by the bundle of
uniform capillaries models. Another feature is the connectivity of the flow channels where bond
aggregation (i.e. how the throats are distributed and arranged) strongly affects yield behavior.

As for the application of numerical methods to \yields\ fluids in porous media, very few studies
can be found on this subject. Moreover, the results, which are reported only for very simple cases
of porous media, cannot be fully assessed. As a result, network modeling seems to be the most
viable candidate for modeling \yields\ fluids in porous media. However, the research in this field
is limited, and hence no final conclusion on the merit of this approach can be reached. In the
followings we present our modeling strategies with a brief assessment of the results and some
general conclusions.

In our attempt to investigate the flow of yield-stress fluids in porous media we use pore-scale
network modeling. Network modeling has been fully described in a number of references (e.g.
\cite{blunt1, blunt2, Sochithesis2007}). Here, we focus only on the part of network modeling
related to yield-stress fluids. The interested reader should refer to those references for other
details. In our \nNEW\ model, yield-stress is described by a \HB\ fluid \cite {skellandbook}:
\begin{equation}\label{}
    \sS = \ysS + C \sR^{n}
    \verb|       | (\sS > \ysS)
\end{equation}
where $\sS$ is the shear stress, $\ysS$ is the \yields\ above which the substance starts flowing,
$C$ is the consistency factor, $\sR$ is the shear rate and $n$ is the flow behavior index. For
\yields\ fluids, the threshold pressure drop above which the flow in a single tube starts is given
by the relation
\begin{equation}\label{yieldCondition}
    \Delta P_{th} = {\frac{2 L \ysS}{R}}
\end{equation}
where $R$ and $L$ are the tube radius and length respectively. For \HB\ fluids, the volumetric flow
rate in a cylindrical capillary at yield is given by
\begin{eqnarray} \label{QHerschel}
    Q = \frac{8\pi}{C^\frac{1}{n}}\left(\frac{L}{\Delta P}\right)^{3}
    \left(\wsS - \ysS\right)^{1+\frac{1}{n}}\left[\frac{\left(\wsS - \ysS\right)^{2}}{3+1/n}+
    \frac{2\ysS \left(\wsS - \ysS\right)}{2+1/n} + \frac{\ysS^{2}}{1+1/n}\right]
    \nonumber \\
    (\wsS > \ysS) \hspace{1.0cm}
\end{eqnarray}
where $\wsS$ ($=\frac{\Delta PR}{2L}$) is the shear stress at the tube wall.

The implementation of yield-stress in the network model is based on the yield condition (i.e.
Equation \ref{yieldCondition}) of its conducting ducts which are assumed to be circular cylinders.
The verification of the yield condition in the individual ducts associates the process of solving
the pressure field in the network. Moreover, when flow occurs the volumetric flow rate of
yield-stress fluids in the network elements is described by the \HB\ relation, i.e. Equation
(\ref{QHerschel}). Another condition is imposed before any duct in the network is allowed to yield,
that is the duct must be part of a non-blocked path spanning the network from the inlet to the
outlet. The reason is that any conducting duct should have a source on one side and a sink on the
other, and this will not happen if either or both sides are blocked.

In our model, the substance before yield is assumed to be fluid with very high but finite viscosity
so the flow virtually vanishes. What justifies this assumption is that the pressure across the
network must communicate. Accordingly, the pressure field in the case of yield-stress fluids is
solved as in the case of non-\yields\ fluids. In both cases, to find the pressure field a set of
simultaneous equations representing the capillaries and satisfying mass conservation have to be
solved subject to the boundary conditions which are the pressures at the inlet and outlet of the
network. This unique and consistent solution, which should mimic the unique physical reality of the
pressure field in the porous medium, is the only mathematically acceptable solution to the problem.
It should be remarked that the assumption of very high but finite zero-stress viscosity for
\yields\ fluids is realistic and supported by experimental evidence.

\begin{figure}[!h]
  \centering{}
  \includegraphics
  [scale=0.5]
  {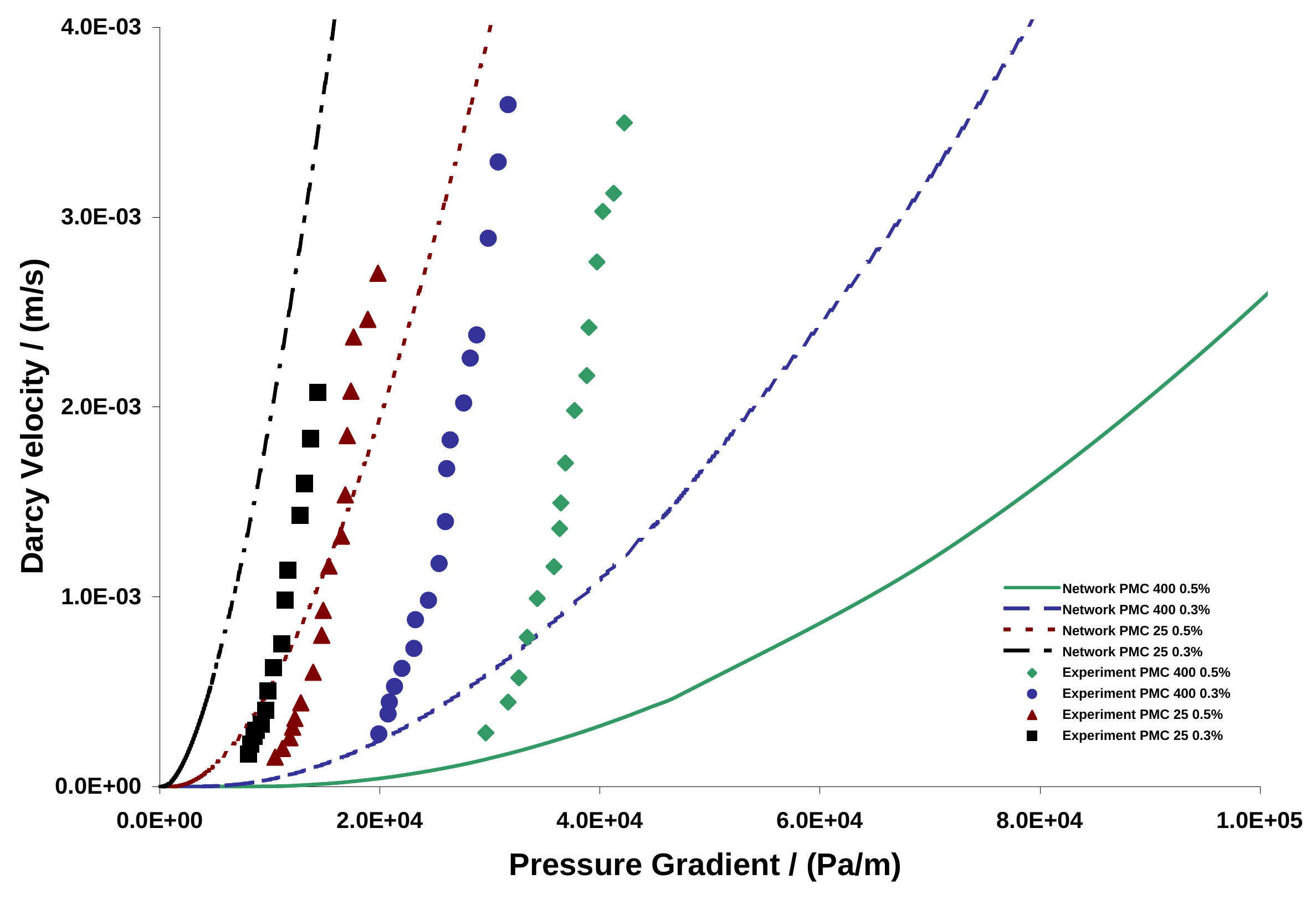}
  \caption[Park experimental data for \HB\ fluid]
  {Park experimental data group for \HB\ aqueous solutions of PMC 400 and PMC 25
  with 0.5\% and 0.3\% weight concentration flowing through a fine packed bed of glass
  beads having $K=366$\,Darcy and $\phi=0.39$ alongside the simulation results obtained with a
  scaled \sandp\ network having same $K$.}
  \label{ParkHBFine}
\end{figure}

\begin{figure}[!h]
  \centering{}
  \includegraphics
  [scale=0.5]
  {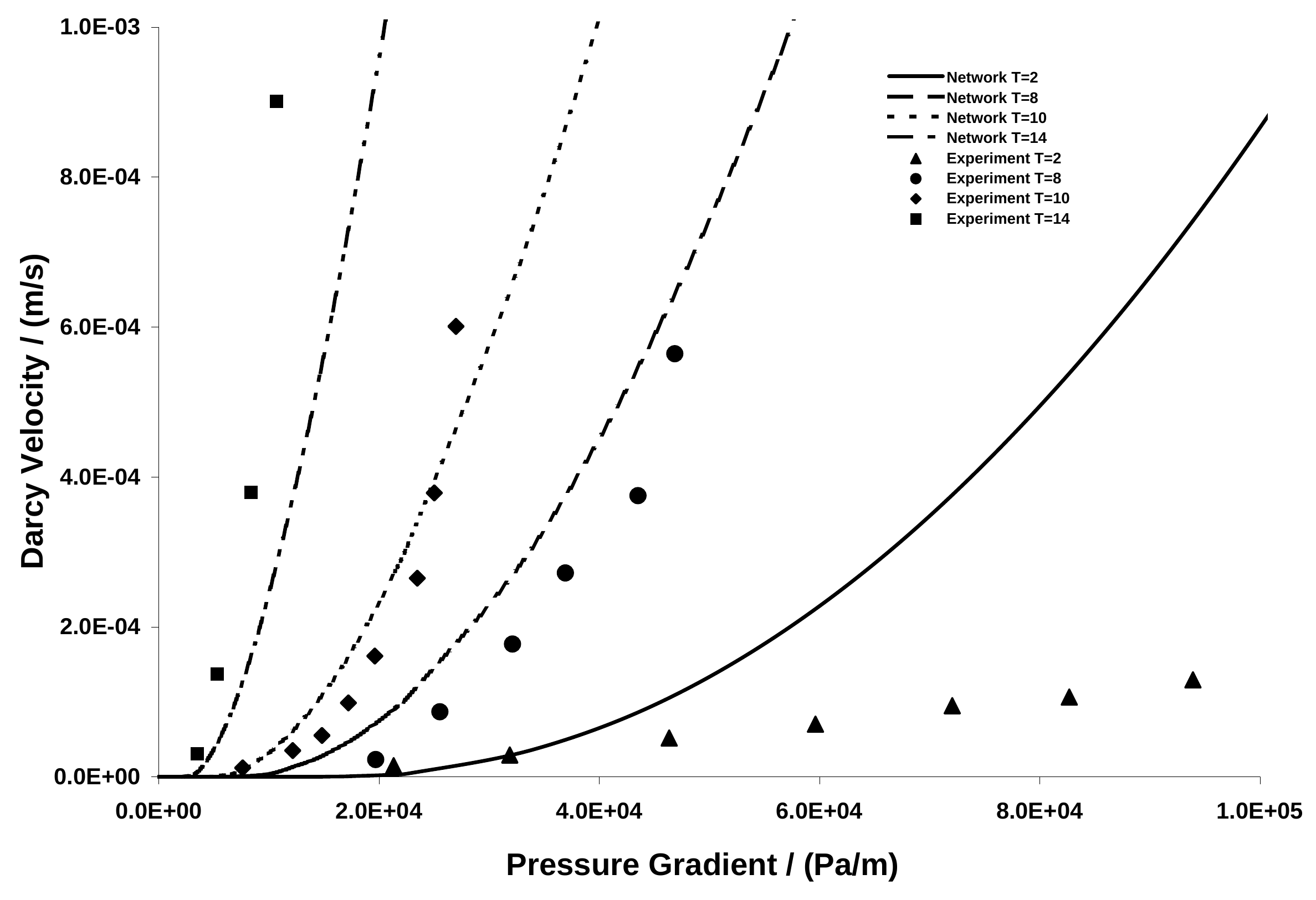}
  \caption[Al-Fariss and Pinder experimental data for \HB\ fluid]
  {Al-Fariss and Pinder experimental data group for \HB\ waxy crude oils flowing
  through a column of sand having $K=1580$\,Darcy and $\phi=0.44$ alongside the simulation results
  obtained with a scaled \sandp\ network having the same $K$. The temperatures, T, are in
  $^{\circ}$C.}
  \label{AlfarissHBCrude}
\end{figure}

\begin{figure}[!h]
  \centering{}
  \includegraphics
  [scale=0.5]
  {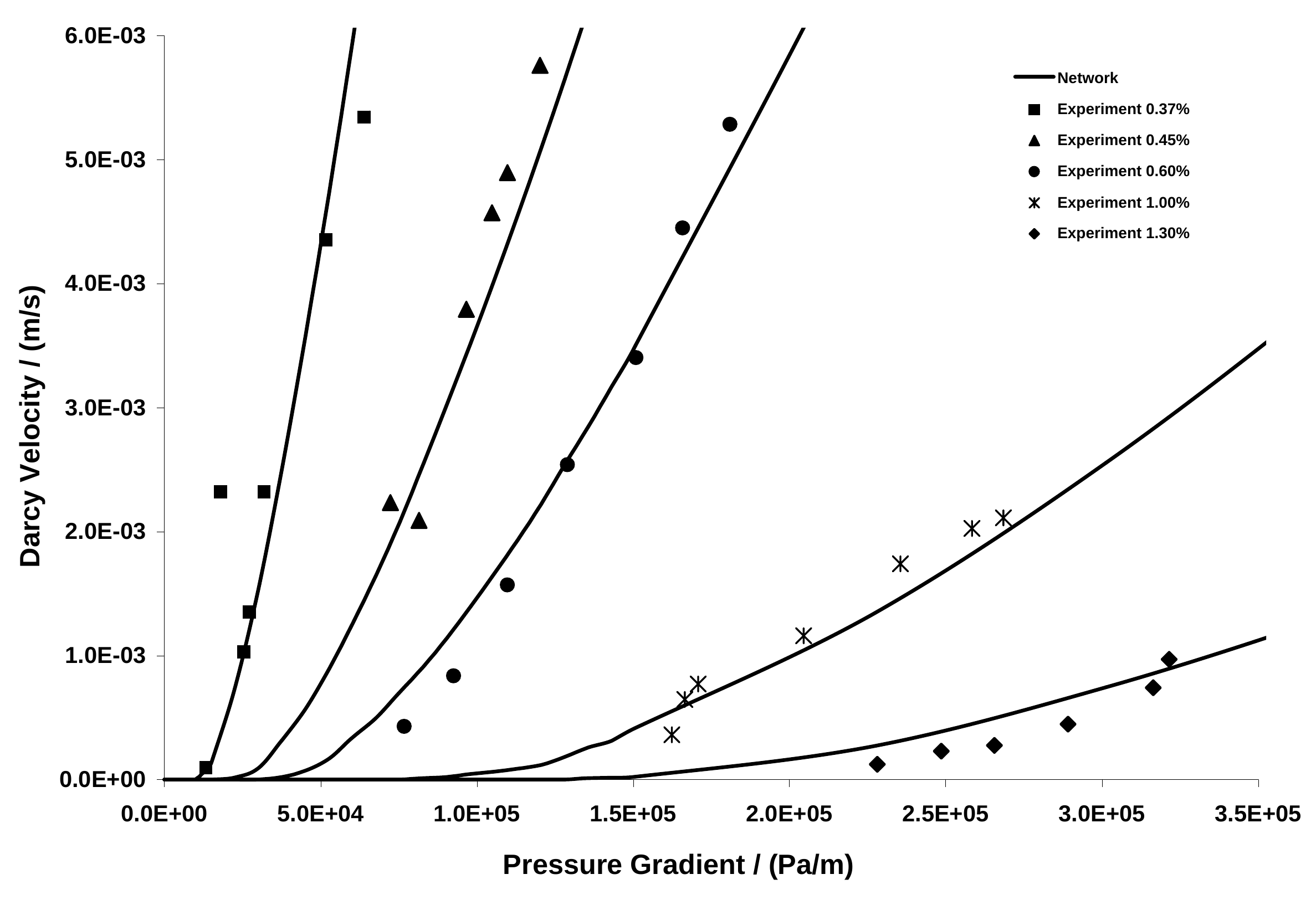}
  \caption[Chase and Dachavijit experimental data for \BING\ fluid]
  {Network simulation results with the corresponding experimental data points
  of Chase and Dachavijit for \BING\ aqueous solutions of Carbopol 941 with various concentrations
  (0.37\%, 0.45\%, 0.60\%, 1.00\% and 1.30\%) flowing through a packed column of glass beads.}
  \label{ChaseHB1}
\end{figure}

The yield-stress flow model, as described above, was implemented in a \nNEW\ computer code
\cite{SochiCode}. Two randomly-distributed networks representing two different porous media, a
\sandp\ and a \Berea\ sandstone \cite{Sochithesis2007}, were used to investigate and assess the
model. Computer-generated cubic networks were also used in this investigation. Theoretical analysis
has been carried out using a simple bundle of capillary tubes of uniform radius. The threshold
yield path was visualized and analyzed to investigate the possibility of aggregation of extreme
elements in a favorable combination that can compromise the reliability of predictions. The
continuity of flow across the network was tested by numerical and visual inspections. Comparison to
experimental data sets found in the literature \cite{parkthesis,alfariss1,chase1} was also carried
out. A representative sample of these data sets with the corresponding simulation results are given
in Figures (\ref{ParkHBFine}, \ref{AlfarissHBCrude}, \ref{ChaseHB1}). As seen, the results are
mixed with obvious failure in some cases.

The theoretical analysis and comparison revealed sensible trends of behavior and sound
implementation. However, as compared to the available experimental data the network model results
for yield-stress fluids are less satisfactory than the results of non-yield-stress fluids. This,
with similar outcome obtained by other researchers, indicates that network modeling in its current
state is not fully capable of describing yield-stress phenomenon. Although the quality of some
experimental data sets is questionable, this reason alone cannot fully explain these failures. One
possible reason is the relative simplicity of the rheological models, such as Bingham, used in
these investigations. These models may be capable of providing a phenomenological description of
\yields\ in simple flow situations. However, they cannot accommodate the underlying physics at pore
level in complex porous media. Consequently, \yields\ as a model parameter obtained in bulk
viscometry may not be appropriate to use in this complex situation.

Another reason is the experimental difficulties associated with \nNEW\ fluids in general. This can
affect the experimental results and introduce complications, such as \vy\ and retention, that may
not be accounted for in the network model. A third reason is the relative simplicity of the current
network modeling approach to \yields\ fluids. This may be supported by the fact that much better
results are normally obtained for non-\yields\ fluids using the same network modeling techniques.
One major limitation of the current network models with regard to \yields\ fluids is the use of
analytical expressions for cylindrical tubes based on the concept of equivalent radius. This is far
from reality where the void space retains highly complex shape and connectivity. Consequently, the
yield condition for cylindrical capillaries becomes invalid approximation to the yield condition
for the intricate flow paths in porous media. The concept of equivalent radius, which is used in
network modeling, though is completely appropriate for \NEW\ fluids and reasonably appropriate for
purely viscous \nNEW\ fluids with no \yields, seems inappropriate for \yields\ fluids as yield
depends on the actual shape of the void space rather than the equivalent radius and flow
conductance. The simplistic nature of the yield condition in porous media is highlighted by the
fact that in almost all cases of disagreement between the network and the experimental results the
network produced a lower yield value.

In summary, \yields\ fluid results are extremely sensitive to how the fluid is characterized, how
the void space is described and how the yield process is modeled. In the absence of a comprehensive
and precise incorporation of all these factors in the network model, pore-scale modeling of
\yields\ fluids in porous media remains a crude approximation that may not produce quantitatively
sensible predictions.

\clearpage
%XXXXXXXXXXXXXXXXXXXXXXXXXXXXXXXXXXXXXXXXXXXXXXXXXXXX
\section{Predicting Threshold Yield Pressure}

Here we discuss the attempts to predict the yield point of a complex porous medium from the void
space description and \yields\ value of an ideal \yields\ fluid without modeling the flow process.
In the literature of \yields\ we can find two well-developed methods proposed for predicting the
yield point of a morphologically-complex network that depicts a porous medium; these are the
Minimum Threshold Path (MTP) and the \percolation\ theory algorithms. In this regard, there is an
implicit assumption that the network is an exact replica of the medium and the \yields\ value
reflects the real \yields\ of the fluid so that any failure of these algorithms can not be
attributed to mismatch or any factor other than flaws in these algorithms. It should be remarked
that the validity of these methods can be tested by experiment.

%XXXXXXXXXXXXXXXXXXXXXXXXXXXXXXXXXXXXXXXXXXXXXXXXXXXX
\subsection{Minimum Threshold Path Algorithms}

Predicting the threshold yield pressure of a \yields\ fluid in porous media in its simplest form
may be regarded as a special case of the more general problem of finding the threshold conducting
path in disordered media that consist of elements with randomly distributed thresholds. This
problem was analyzed by Roux and Hansen \cite {roux1} in the context of studying the conduction of
an electric network of diodes by considering two different cases, one in which the path is directed
(no backtracking) and one in which it is not. They suggested that the minimum overall threshold
potential difference across the network is akin to a \percolation\ threshold and investigated its
dependence on the lattice size. Kharabaf and Yortsos \cite {kharabaf1} noticed that a firm
connection of the lattice-threshold problem to \percolation\ appears to be lacking and the relation
of the minimum threshold path to the minimum path of \percolation, if it indeed exists, is not
self-evident. They presented a new algorithm, \InvPM\ (IPM), for the construction of the MTP to
study its properties.  The \InvPM\ was further extended by Chen \etal\ \cite {chen3} to incorporate
dynamic effects due to viscous friction following the onset of mobilization. In the course of our
investigation, another MTP algorithm called Path of Minimum Pressure (PMP), which is
computationally more efficient than IPM, was developed by the author. Both algorithms were
implemented in the \nNEW\ code. The results of these algorithms, which are identical in most cases,
were analyzed and compared to the experimental data and the network flow simulation predictions. In
the following we outline and assess these algorithms.

\begin{figure}[!h]
  \centering{}
  \includegraphics
  [scale=0.90, trim = 0 -20 0 20]
  {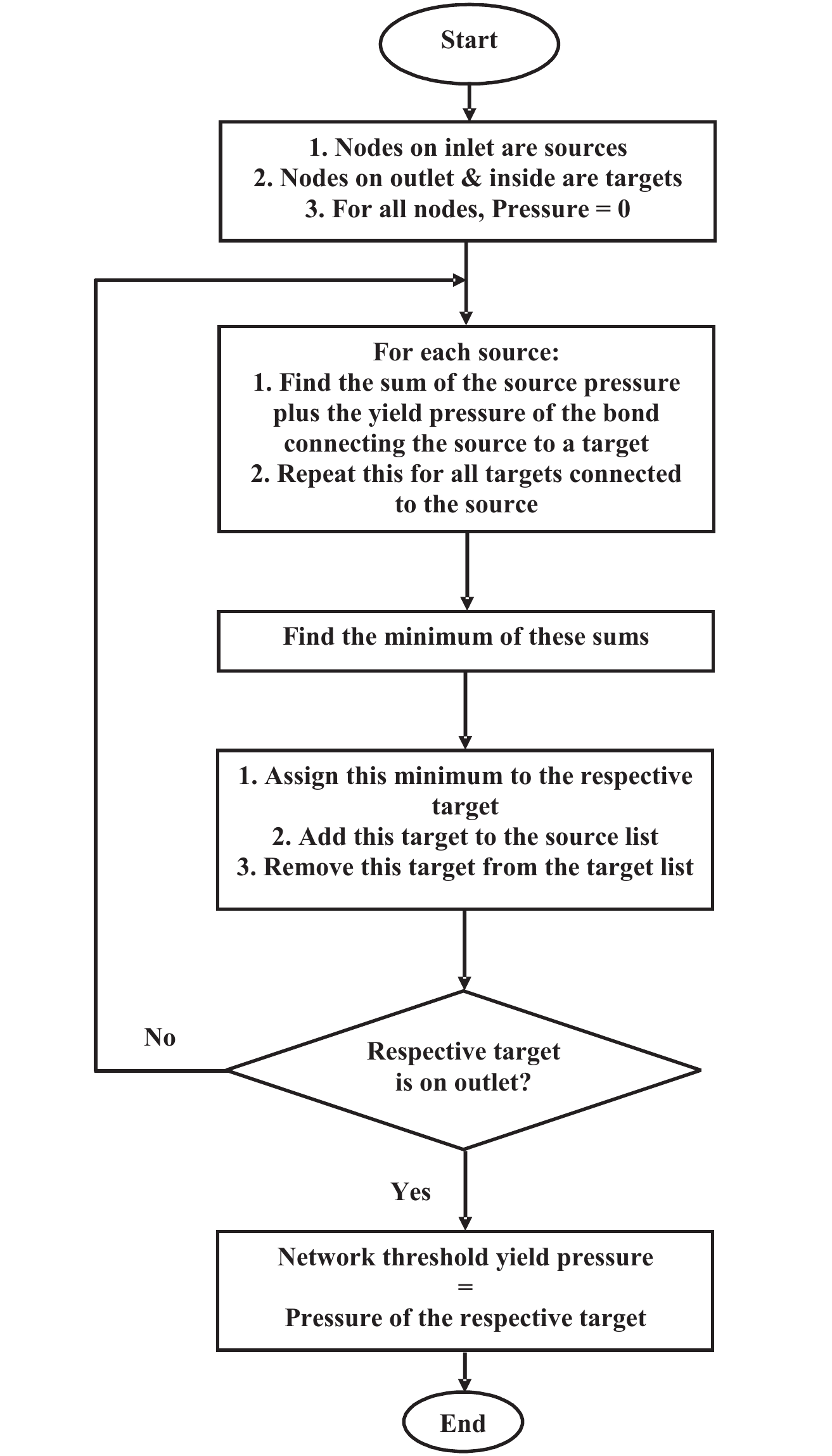}
  \caption[Flowchart of the \InvPM\ (IPM) algorithm]
  {Flowchart of the \InvPM\ (IPM) algorithm.}
  \label{IPMFlowchart}
\end{figure}

The IPM is a way for finding the inlet-to-outlet path that minimizes the sum of the values of a
property assigned to the individual elements of the network, and hence finding this minimum. For a
\yields\ fluid, this reduces to finding the inlet-to-outlet path that minimizes the yield pressure.
The yield pressure of this path is taken as the network threshold yield pressure. An algorithm to
find the threshold yield pressure according to IPM is outlined below:
\begin{enumerate}

\item Initially, the nodes on the inlet are considered to be sources and the nodes on the outlet and
inside are targets. The inlet nodes are assigned a pressure value of 0.0. According to the IPM, a
source cannot be a target and vice versa, i.e. they are disjoint sets and remain so in all stages.

\item Starting from the source nodes, a step forward is made in which the yield front advances one bond
from a single source node. The condition for choosing this step is that the sum of the source
pressure plus the yield pressure of the bond connecting the source to the target node is the
minimum of all similar sums from the source nodes to the possible target nodes. This sum is
assigned to the target node.

\item This target node loses its status as a target and obtains the status of a source.

\item The last two steps are repeated until the outlet is reached, i.e. when the target is an outlet
node. The pressure assigned to this target node is regarded as the threshold yield pressure of the
network.

\end{enumerate}
A flowchart of the IPM algorithm is given in Figure (\ref{IPMFlowchart}).

\begin{figure}[!h]
  \centering{}
  \includegraphics
  [scale=0.9, trim = 0 -5 0 20]
  {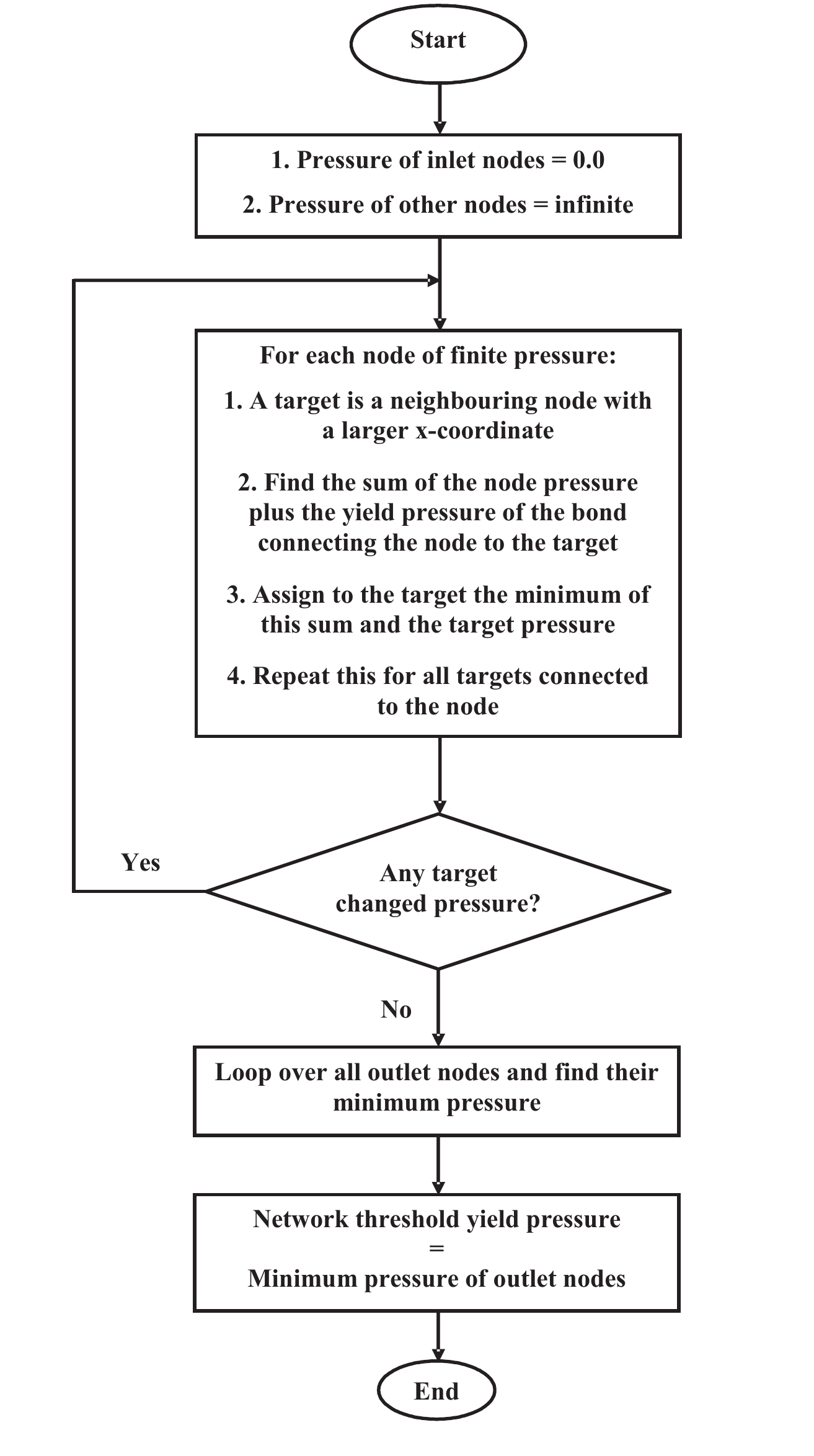}
  \caption[Flowchart of the \PathMP\ (PMP) algorithm]
  {Flowchart of the \PathMP\ (PMP) algorithm.}
  \label{PMPFlowchart}
\end{figure}

The PMP is based on a similar assumption to that upon which the IPM is based, that is the network
threshold yield pressure is the minimum sum of the threshold yield pressures of the individual
elements of all possible paths from the inlet to the outlet. However, it is computationally
different and is more efficient than the IPM in terms of the CPU time and memory. According to the
PMP, to find the threshold yield pressure of a network all possible paths of serially-connected
bonds from inlet to outlet are inspected. For each path, the threshold yield pressure of each bond
is computed and the sum of these pressures is found. The network threshold yield pressure is then
taken as the minimum of these sums. The implementation of this algorithm can be outlined as
follows:
\begin{enumerate}

\item The nodes on the inlet are initialized with zero pressure while the remaining nodes are initialized
with infinite pressure (very high value).

\item A source is an inlet or an inside node with a finite pressure while a target is a node connected to
a source through a single bond in the direction of pressure gradient.

\item Looping over all sources, the threshold yield pressure of each bond connecting a source to a target
is computed and the sum of this yield pressure plus the pressure of the source node is found. The
minimum of this sum and the current yield pressure of the target node is assigned to the target
node.

\item The loop in the last step is repeated until a stable state is reached when no target node changes
its pressure during looping over all sources.

\item A loop over all outlet nodes is used to find the minimum pressure assigned to the outlet nodes.
This pressure is taken as the network threshold yield pressure.

\end{enumerate}
The PMP algorithm is graphically depicted in a flow chart in Figure (\ref{PMPFlowchart}).

As implemented in the \nNEW\ code, the IPM and PMP produce very similar results. However, they both
predict lower threshold yield pressure compared to the network flow simulation results and the
experimental data.

There are two possibilities for defining yield stress fluids before yield: either solid-like
substances or liquids with very high viscosity. According to the first, the most sensible way for
modeling a presumed pressure gradient inside a medium before yield is to be a constant, that is the
pressure drop across the medium is a linear function of the spatial coordinate in the flow
direction. The reason is that any other assumption requires justification. In the second case, the
fluid should be treated like non-\yields\ fluids and hence the pressure field inside the porous
medium should be subject to the consistency criterion given in \S\ \ref{ModelingYS}. The logic is
that the magnitude of the viscosity should have no effect on the flow behavior as long as the
material is assumed to be fluid.

Several arguments can be presented against the MTP algorithms for predicting the yield point of a
network representing a medium. Although some arguments may be more obvious for a network with
cylindrical ducts, they are valid in general for regular and irregular geometries of flow channels.
Some of these arguments are outlined below

\begin{itemize}

\item The MTP algorithms are based on the assumption that the threshold yield pressure (TYP) of an
ensemble of serially connected bonds is the sum of their yield pressures. This assumption can be
challenged by the fact that for a non-uniform ensemble (i.e. an ensemble whose elements have
different TYPs) the pressure gradient across the ensemble should reach the threshold yield gradient
of the bottleneck (i.e. the element with the highest TYP) of the ensemble if yield should occur.
Consequently, the TYP of the ensemble will be higher than the sum of the TYPs of the individual
elements. This argument may be more obvious if \yields\ fluids are regarded as solids and a linear
pressure drop is assumed.

\item Assuming that \yields\ substances before yield are fluids with high viscosity, the dynamic aspects
of the pressure field are neglected because the aim of the MTP algorithms is to find a collection
of bonds inside the network with a certain condition based on the intrinsic properties of these
elements irrespective of the pressure field. The reality is that the bonds are part of a network
that is subject to a pressure field, so the pressure across each individual element must comply
with a dynamically stable and mathematically consistent pressure configuration over the whole
network. The MTP algorithms rely for justification on a hidden assumption that the minimum sum
condition is identical or equivalent to the stable configuration condition. This assumption is
disputable because it is very unlikely that a stable configuration will require a pressure drop
across each element of the minimum threshold path that is identical to the threshold yield pressure
of that element. Therefor, it is not clear that the actual path of yield must coincide, totally or
partially, with the path of the MTP algorithms let alone that the actual value of yield pressure
must be predicted by these algorithms.

\item The MTP algorithms that allow backtracking have another drawback that is in some cases the minimum
threshold path requires a physically unacceptable pressure configuration. This is more obvious if
the \yields\ substances are assumed to be solid before yield.

\item The effect of tortuosity is ignored in the MTP algorithms since they implicitly assume that the
path of yield is an ensemble of serially-connected and linearly-aligned tubes, whereas in reality
the path is generally tortuous as it is part of a network and can communicate with the global
pressure field through the intermediate nodes. The effect of tortuosity, which is more obvious for
the solid assumption, is a possible increase in the external threshold pressure gradient and a
possible change in the bottleneck.

\end{itemize}

%XXXXXXXXXXXXXXXXXXXXXXXXXXXXXXXXXXXXXXXXXXXXXXXXXXXX
\subsection{Percolation Theory}

Concerning the \percolation\ approach, it is tempting to consider the conduct of \yields\ fluids in
porous media as a \percolation\ phenomenon to be analyzed by the classical \percolation\ theory.
However, three reasons, at least, may suggest otherwise

\begin{enumerate}

\item The conventional \percolation\ models can be applied only if the conducting elements are
homogeneous, i.e. it is assumed in these models that the intrinsic property of all elements in the
network are equal. However, this assumption cannot be justified for most kinds of media where the
elements vary in their conduction and yield properties. Therefore, to apply \percolation\
principles, a generalization to the conventional \percolation\ theory is needed as suggested by
Selyakov and Kadet \cite {selyakovbook}.

\item The network elements cannot yield independently as a spanning path bridging the inlet to the outlet
is a necessary condition for yield. This contradicts the \percolation\ theory assumption that the
elements conduct independently.

\item The pure \percolation\ approach ignores the dynamic aspects of the pressure field, that is a stable
pressure configuration is a necessary condition which may not coincide with the simple
\percolation\ requirement. The \percolation\ condition, as required by the classic \percolation\
theory, determines the stage at which the network starts flowing according to the intrinsic
property of its elements as an ensemble of conducting bonds regardless of the dynamic aspects
introduced by the external pressure field, as explained earlier.

\end{enumerate}

In a series of studies on generation and mobilization of foams in porous media, Rossen \etal\
\cite{rossen1, rossen2} analyzed the threshold yield pressure using \percolation\ theory concepts
and suggested a simple \percolation\ model. In this model, the \percolation\ cluster is first
found, then the minimum threshold path was approximated as a subset of this cluster that samples
those bonds with the smallest individual thresholds \cite {chen3}. This approach relies on the
validity of applying \percolation\ theory to \yields, which is disputed. Moreover, it is a mere
coincidence if the yield path is contained within the \percolation\ sample. Yield is an on/off
process which critically depends on factors other than smallness of individual thresholds. These
factors include the particular distribution and configuration of these elements, being within a
larger network and hence being able to communicate with the global pressure field, and the dynamic
aspects of the pressure field and stability requirement. Any approximation, therefore, has little
meaning in this context.

\newpage
%XXXXXXXXXXXXXXXXXXXXXXXXXXXXXXXXXXXXXXXXXXXXXXXXXXXXXXXXXXXXXXXXX
\section{Conclusions and Discussions} \label{Conclusions}

In this study, we developed and implemented a \nNEW\ flow simulation model for the flow of
yield-stress fluids in porous media, and compared the simulation results to a number of
experimental data sets found in the literature. We also investigated two minimum threshold path
algorithms and the percolation theory as a basis for predicting the threshold yield pressure of a
network depicting a porous medium. General results and conclusions that can be drawn from this
study are

\begin{itemize}

\item In several cases, the simulation results as compared to the experimental data are not
satisfactory. One possible reason is inadequate representation of the pore space structure with
regard to the yield process, i.e. the yield in a network depends on the actual shape of the void
space rather than the flow conductance of the pores and throats. Other reasons include experimental
errors and the occurrence of other physical phenomena which are not accounted for in our model such
as precipitation and adsorption.

\item The analysis of minimum threshold path algorithms (i.e. IPM and PMP) revealed that these
algorithms are too simplistic and hence cannot produce reliable predictions for the threshold yield
pressure of a network. Compared to the experimental data and the network simulation results, these
algorithms predict lower threshold yield pressure.

\item The assessment of the percolation theory suggested that percolation may not be suitable for
modeling yield-stress in porous media.

\end{itemize}

The final conclusion is that the current flow modeling methodologies cannot cope with the
complexity of yield-stress fluids in porous media. More elaborate strategies are required to make
progress on this front.

\newpage
%XXXXXXXXXXXXXXXXXXXXXXXXXXXXXXXXXXXXXXXXXXXXXXXXXXXXXXXXXXXXXXXXXXX
\phantomsection \addcontentsline{toc}{section}{Nomenclature} %
{\noindent \LARGE \bf Nomenclature} \vspace{0.5cm}

\begin{supertabular}{ll}
  $\sR$                 & strain rate (s$^{-1}$) \\
  $\sS$                 & stress (Pa) \\
  $\ysS$                & \yields\ (Pa) \\
  $\wsS$                & stress at tube wall (Pa) \\
  $\phi$                & porosity \\
\\
  $C$                   & consistency factor in \HB\ model (Pa.s$^{n}$) \\
  $K$                   & absolute permeability (m$^{2}$) \\
  $L$                   & tube length (m) \\
  $n$                   & flow behavior index \\
  $P$                   & pressure (Pa) \\
  $\Delta P$            & pressure drop (Pa) \\
  $\Delta P_{th}$       & threshold pressure drop (Pa) \\
  $Q$                   & volumetric flow rate (m$^{3}$.s$^{-1}$) \\
  $R$                   & tube radius (m) \\
  T                     & temperature (K, $^{\circ}$C) \\
\\
  IPM                   & \InvPM\ algorithm \\
  MTP                   & Minimum Threshold Path \\
  PMP                   & \PathMP\ algorithm \\
  TYP                   & Threshold Yield Pressure \\
\end{supertabular}

\newpage
\phantomsection \addcontentsline{toc}{section}{References} %
\bibliographystyle{unsrt}
\bibliography{Biblio}

\end{document}